\documentclass[10pt,a4paper,onecolumn]{article}
\usepackage{marginnote}
\usepackage{graphicx}
\usepackage{xcolor}
\usepackage{authblk,etoolbox}
\usepackage{titlesec}
\usepackage{calc}
\usepackage{tikz}
\usepackage{hyperref}
\hypersetup{colorlinks,breaklinks=true,
            urlcolor=[rgb]{0.0, 0.5, 1.0},
            linkcolor=[rgb]{0.0, 0.5, 1.0}}
\usepackage{caption}
\usepackage{tcolorbox}
\usepackage{amssymb,amsmath}
\usepackage{ifxetex,ifluatex}
\usepackage{seqsplit}
\usepackage{xstring}

\usepackage{float}
\let\origfigure\figure
\let\endorigfigure\endfigure
\renewenvironment{figure}[1][2] {
    \expandafter\origfigure\expandafter[H]
} {
    \endorigfigure
}

\usepackage{fixltx2e} 
\usepackage[
  backend=biber,
]{biblatex}
\bibliography{paper.bib}

\let\textttOrig=\texttt
\def\texttt#1{\expandafter\textttOrig{\seqsplit{#1}}}
\renewcommand{\seqinsert}{\ifmmode
  \allowbreak
  \else\penalty6000\hspace{0pt plus 0.02em}\fi}

\makeatletter
\let\href@Orig=\href
\def\href@Urllike#1#2{\href@Orig{#1}{\begingroup
    \def\Url@String{#2}\Url@FormatString
    \endgroup}}
\def\href@Notdoi#1#2{\def\tempa{#1}\def\tempb{#2}%
  \ifx\tempa\tempb\relax\href@Urllike{#1}{#2}\else
  \href@Orig{#1}{#2}\fi}
\def\href#1#2{%
  \IfBeginWith{#1}{https://doi.org}%
  {\href@Urllike{#1}{#2}}{\href@Notdoi{#1}{#2}}}
\makeatother

\newlength{\cslhangindent}
\newlength{\csllabelwidth}
\newenvironment{CSLReferences}[3] 
 {
  \setlength{\parindent}{0pt}
  \ifodd #1 \everypar{\setlength{\hangindent}{\cslhangindent}}\ignorespaces\fi
  \ifnum #2 > 0
  \setlength{\parskip}{#2\baselineskip}
  \fi
 }%
 {}
\usepackage{calc}

\usepackage[top=3.5cm, bottom=3cm, right=1.5cm, left=1.0cm,
            headheight=2.2cm, reversemp, includemp, marginparwidth=4.5cm]{geometry}

\titleformat{\section}
  {\normalfont\sffamily\Large\bfseries}
  {}{0pt}{}
\titleformat{\subsection}
  {\normalfont\sffamily\large\bfseries}
  {}{0pt}{}
\titleformat{\subsubsection}
  {\normalfont\sffamily\bfseries}
  {}{0pt}{}
\titleformat*{\paragraph}
  {\sffamily\normalsize}

\usepackage{fancyhdr}
\makeatletter
\let\ps@plain\ps@fancy
\fancyheadoffset[L]{4.5cm}
\fancyfootoffset[L]{4.5cm}

\definecolor{linky}{rgb}{0.0, 0.5, 1.0}

\newtcolorbox{repobox}
   {colback=red, colframe=red!75!black,
     boxrule=0.5pt, arc=2pt, left=6pt, right=6pt, top=3pt, bottom=3pt}

\patchcmd{\@maketitle}{center}{flushleft}{}{}
\patchcmd{\@maketitle}{center}{flushleft}{}{}
\patchcmd{\@maketitle}{\LARGE}{\LARGE\sffamily}{}{}
\def\maketitle{{%
  
  \AB@maketitle}}
\makeatletter
\renewcommand\AB@affilsepx{ \protect\Affilfont}
\renewcommand\AB@affilnote[1]{{\bfseries #1}\hspace{3pt}}
\renewcommand{\affil}[2][]%
   {\newaffiltrue\let\AB@blk@and\AB@pand
      \if\relax#1\relax\def\AB@note{\AB@thenote}\else\def\AB@note{#1}%
        \setcounter{Maxaffil}{0}\fi
        \begingroup
        \let\href=\href@Orig
        \let\texttt=\textttOrig
        \let\protect\@unexpandable@protect
        \def\thanks{\protect\thanks}\def\footnote{\protect\footnote}%
        \@temptokena=\expandafter{\AB@authors}%
        {\def\\{\protect\\\protect\Affilfont}\xdef\AB@temp{#2}}%
         \xdef\AB@authors{\the\@temptokena\AB@las\AB@au@str
         \protect\\[\affilsep]\protect\Affilfont\AB@temp}%
         \gdef\AB@las{}\gdef\AB@au@str{}%
        {\def\\{, \ignorespaces}\xdef\AB@temp{#2}}%
        \@temptokena=\expandafter{\AB@affillist}%
        \xdef\AB@affillist{\the\@temptokena \AB@affilsep
          \AB@affilnote{\AB@note}\protect\Affilfont\AB@temp}%
      \endgroup
       \let\AB@affilsep\AB@affilsepx
}
\makeatother

\renewcommand\Affilfont{\sffamily\small\mdseries}
\ifnum 0\ifxetex 1\fi\ifluatex 1\fi=0 
  \usepackage[T1]{fontenc}
  \usepackage[utf8]{inputenc}

\else 
  \ifxetex
    \usepackage{mathspec}
    \usepackage{fontspec}

  \else
    \usepackage{fontspec}
  \fi
  \defaultfontfeatures{Ligatures=TeX,Scale=MatchLowercase}

\fi
\IfFileExists{upquote.sty}{\usepackage{upquote}}{}
\IfFileExists{microtype.sty}{%
\usepackage{microtype}
\UseMicrotypeSet[protrusion]{basicmath} 
}{}

\usepackage{hyperref}
\hypersetup{unicode=true,
            pdftitle={Danna-Sep: Unite to separate them all},
            pdfborder={0 0 0},
            breaklinks=true}
\usepackage{longtable,booktabs}

\let\addcontentslineOrig=\addcontentsline
\def\addcontentsline#1#2#3{\bgroup
  \let\texttt=\textttOrig\addcontentslineOrig{#1}{#2}{#3}\egroup}
\let\markbothOrig\markboth
\def\markboth#1#2{\bgroup
  \let\texttt=\textttOrig\markbothOrig{#1}{#2}\egroup}
\let\markrightOrig\markright
\def\markright#1{\bgroup
  \let\texttt=\textttOrig\markrightOrig{#1}\egroup}

\usepackage{graphicx,grffile}
\makeatletter
\def\maxwidth{\ifdim\Gin@nat@width>\linewidth\linewidth\else\Gin@nat@width\fi}
\def\maxheight{\ifdim\Gin@nat@height>\textheight\textheight\else\Gin@nat@height\fi}
\makeatother
\setkeys{Gin}{width=\maxwidth,height=\maxheight,keepaspectratio}
\IfFileExists{parskip.sty}{%
\usepackage{parskip}
}{
\setlength{\parindent}{0pt}
\setlength{\parskip}{6pt plus 2pt minus 1pt}
}
\ifx\paragraph\undefined\else
\let\oldparagraph\paragraph
\renewcommand{\paragraph}[1]{\oldparagraph{#1}\mbox{}}
\fi
\ifx\subparagraph\undefined\else
\let\oldsubparagraph\subparagraph
\renewcommand{\subparagraph}[1]{\oldsubparagraph{#1}\mbox{}}
\fi

\title{Danna-Sep: Unite to separate them all}

        \author[1]{Chin-Yun Yu}
          \author[2]{Kin-Wai Cheuk}
    
      \affil[1]{Independent Researcher}
      \affil[2]{Information Systems and Technology Design, Singapore
University of Technology and Design}
  \date{\vspace{-7ex}}

\begin{document}
\maketitle

\marginpar{

  \begin{flushleft}
  \sffamily\small

  \vspace{2mm}

  \par\noindent\hrulefill\par

  \vspace{2mm}

  \vspace{2mm}
  {\bfseries License}\\
  Authors of papers retain copyright and release the work under a Creative Commons Attribution 4.0 International License (\href{http://creativecommons.org/licenses/by/4.0/}{\color{linky}{CC BY 4.0}}).

  \vspace{4mm}
  {\bfseries In partnership with}\\
  \vspace{2mm}
  \includegraphics[width=4cm]{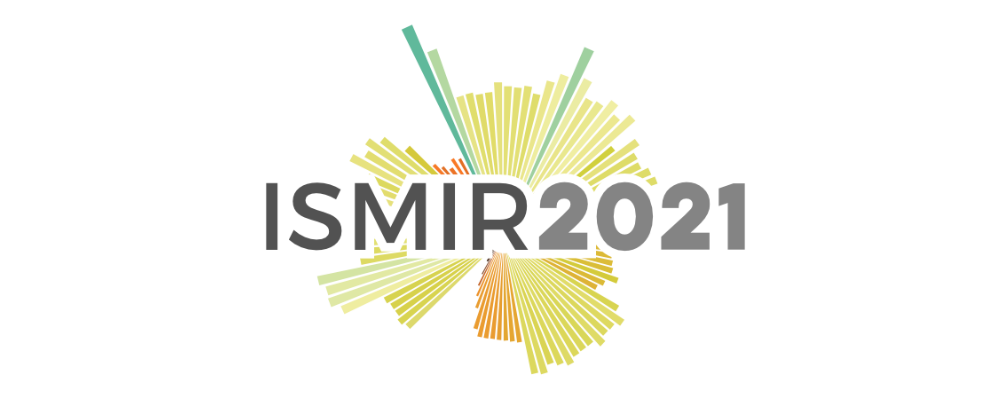}

  \end{flushleft}
}

\hypertarget{abstract}{%
\section{Abstract}\label{abstract}}

Deep learning-based music source separation has gained a lot of interest
in the last decades. Most of the existing methods operate with either
spectrograms or waveforms. Spectrogram-based models learn suitable masks
for separating magnitude spectrogram into different sources, and
waveform-based models directly generate waveforms of individual sources.
The two types of models have complementary strengths; the former is
superior given harmonic sources such as vocals, while the latter
demonstrates better results for percussion and bass instruments. In this
work, we improved upon the state-of-the-art (SoTA) models and
successfully combined the best of both worlds. The backbones of the
proposed framework, dubbed Danna-Sep\footnote{\url{https://github.com/yoyololicon/danna-sep}},
are two spectrogram-based models including a modified X-UMX and U-Net,
and an enhanced Demucs as the waveform-based model. Given an input of
mixture, we linearly combined respective outputs from the three models
to obtain the final result. We showed in the experiments that, despite
its simplicity, Danna-Sep surpassed the SoTA models by a large margin in
terms of Source-to-Distortion Ratio.

\hypertarget{method}{%
\section{Method}\label{method}}

Danna-Sep is a combination of three different models: X-UMX, U-Net, and
Demucs. We describe the enhancements made for each model in the
following subsections.

\begin{figure}
\centering
\includegraphics[width=1\textwidth,height=\textheight]{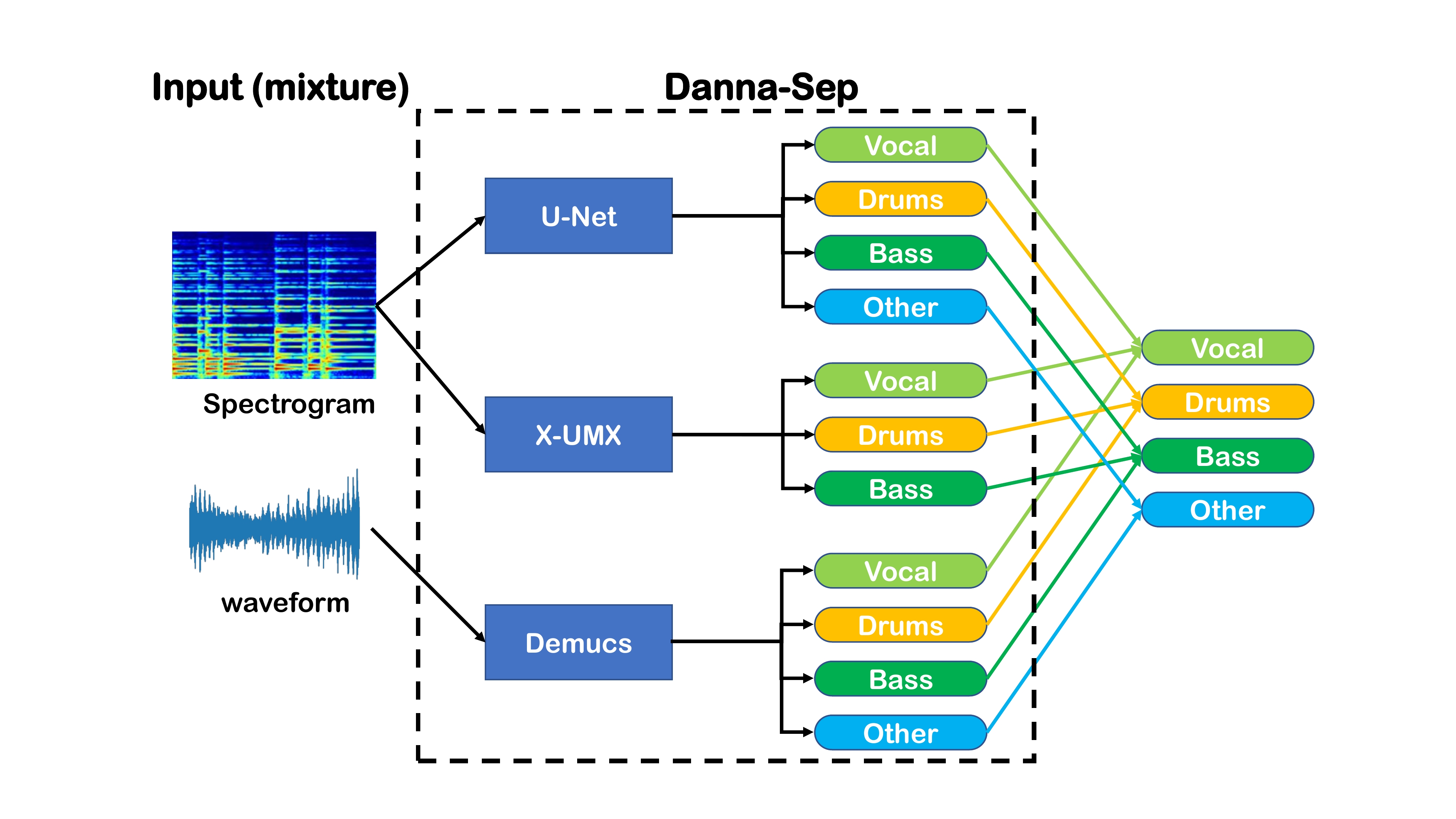}
\caption{The schematic diagram of our proposed system.}
\end{figure}

\hypertarget{x-umx}{%
\subsection{X-UMX}\label{x-umx}}

X-UMX (Sawata et al., 2020) improved upon UMX(St "oter et al., 2019) by
concatenating hidden layers of UMX to enable sharing information among
all target instruments. We trained the model using the same time-domain
loss as the original X-UMX, but modified the frequency-domain loss for
\(J\) sources as follows:

\[\mathcal{L}_{MSE}^J = \sum_{j=1}^J\sum_{t,f}|Y_j(t, f) - \hat{Y}_j(t, f)|^2\]

where \(Y_j(t, f)\) and \(\hat{Y}_j(t, f)\) are ground-truth and
estimated time-frequency representations for the \(j\)-th source,
respectively. That is, instead of taking norm of the absolute value as
in the original X-UMX, we calculated Euclidean norm in the complex
domain. Also, we incorporated Multichannel Wiener Filtering
(MWF)(Liutkus \& Stöter, 2019) into our training pipeline in order to
train our model in an end-to-end fashion. We initialized our modified
X-UMX with the official pre-trained X-UMX weights\footnote{\url{https://zenodo.org/record/4740378/files/pretrained_xumx_musdb18HQ.pth}}
and continued training for approximately 70 epochs with a batch size of
four.

\hypertarget{u-net}{%
\subsection{U-Net}\label{u-net}}

The encoder and decoder of our U-Net consist of six D3 Blocks (Takahashi
\& Mitsufuji, 2021) and we added two layers of 2D local attention
(Parmar et al., 2018) layers at the bottleneck. We used the same loss
function as X-UMX during training but with MWF being disabled. The
approximated training time was nine days with a batch size of 16 on four
Tesla V100 GPUs. We also experimented with using biaxial biLSTM along
the time and frequency axes as the bottleneck layers, but it took
slightly longer to train yet offered a negligible improvement.

\hypertarget{demucs}{%
\subsection{Demucs}\label{demucs}}

For Demucs (Défossez et al., 2019), we built upon the variant with 48
hidden channels, and enhanced the model by replacing the decoder with
four independent decoders responsible for four respective sources. Each
decoder has the same architecture as the original decoder, except for
size of the hidden channel which was reduced to 24. This makes the total
number of parameters comparable with the original Demucs. The training
loss aggregates the L1-norm between estimated and ground-truth waveforms
of the four sources. The model took approximately 10 days to train on a
single RTX 3070 using mixed precision with a batch size of 16, and four
steps of gradient accumulation.

\hypertarget{danna-sep}{%
\subsection{Danna-Sep}\label{danna-sep}}

In order to obtain the final output of our framework, we calculated
weighted average of individual outputs from the above-mentioned models.
Experiments were conducted to search for optimal weighting. The optimal
weights for each source, types of input domain (T for waveforms, TF for
frequency masking), and the sizes of the models are given in the
following table.

\begin{longtable}[]{@{}lcccccc@{}}
\toprule
& Drums & Bass & Other & Vocals & Input Domain & Size
(Mb)\tabularnewline
\midrule
\endhead
X-UMX & 0.2 & 0.1 & 0 & 0.2 & TF & 136\tabularnewline
U-Net & 0.2 & 0.17 & 0.5 & 0.4 & TF & 61\tabularnewline
Demucs & 0.6 & 0.73 & 0.5 & 0.4 & T & 733\tabularnewline
\bottomrule
\end{longtable}

All models were trained on the training set of musdb18-hq (Rafii et al.,
2019) using an Adam optmizier(Kingma \& Ba, 2014).

\hypertarget{separation-performances}{%
\section{Separation performances}\label{separation-performances}}

For a fair comparison, we trained all the models with musdb18-hq (Rafii
et al., 2019) and performed the evaluation using the compressed version
of the dataset (Rafii et al., 2017). One iteration of MWF was used for
X-UMX and U-Net, and we didn't apply the shift trick (Défossez et al.,
2019) for our enhanced Demucs. In the table below, we report the
Signal-to-Distortion Ratio (SDR) (Vincent et al., 2006), calculated
using \emph{museval} (Stöter \& Liutkus, 2019), attained by our modified
models and the original counterparts, as well as the proposed framework.

\begin{longtable}[]{@{}lccccc@{}}
\toprule
& Drums & Bass & Other & Vocals & Avg.\tabularnewline
\midrule
\endhead
X-UMX (baseline) & 6.44 & 5.54 & 4.46 & 6.54 & 5.75\tabularnewline
X-UMX (ours) & 6.71 & 5.79 & 4.63 & 6.93 & 6.02\tabularnewline
U-Net (ours) & 6.43 & 5.35 & 4.67 & 7.05 & 5.87\tabularnewline
Demucs (baseline) & 6.67 & 6.98 & 4.33 & 6.89 & 6.21\tabularnewline
Demucs (ours) & 6.72 & 6.97 & 4.4 & 6.88 & 6.24\tabularnewline
Danna-Sep & \textbf{7.2} & \textbf{7.05} & \textbf{5.2} & \textbf{7.63}
& \textbf{6.77}\tabularnewline
\bottomrule
\end{longtable}

As can be seen from the table, our modified X-UMX gained an extra 0.27
dB on average SDR compared to the original X-UMX. The enhanced Demucs
outperformed the original model by 0.03 dB of SDR, despite the fact that
the shift trick was not applied. Notably, Danna-Sep surpassed both the
original and enhanced Demucs by a large margin (+0.53 dB on average
SDR). Altogether, the results demonstrate the efficacy of the proposed
fusion method in addition to our modifications to the training scheme
and architecture. The proposed framework, however, is more reliant on
computing power due to the nature of model fusion, which we would like
to address in furture work.

\hypertarget{acknowledgements}{%
\section{Acknowledgements}\label{acknowledgements}}

We acknowledge contributions from Sung-Lin Yeh and Yu-Te Wu, and
supports from Yin-Jyun Luo and Showmin Wang during the genesis of this
project.

\hypertarget{references}{%
\section*{References}\label{references}}
\addcontentsline{toc}{section}{References}

\hypertarget{refs}{}
\begin{CSLReferences}{1}{0}
\leavevmode\hypertarget{ref-defossez2019music}{}%
Défossez, A., Usunier, N., Bottou, L., \& Bach, F. (2019). Music source
separation in the waveform domain. \emph{arXiv Preprint
arXiv:1911.13254}.

\leavevmode\hypertarget{ref-kingma2014adam}{}%
Kingma, D. P., \& Ba, J. (2014). Adam: A method for stochastic
optimization. \emph{arXiv Preprint arXiv:1412.6980}.

\leavevmode\hypertarget{ref-antoine_liutkus_2019_3269749}{}%
Liutkus, A., \& Stöter, F.-R. (2019). \emph{Sigsep/norbert: First
official norbert release} (Version v0.2.0) {[}Computer software{]}.
Zenodo. \url{https://doi.org/10.5281/zenodo.3269749}

\leavevmode\hypertarget{ref-parmar2018image}{}%
Parmar, N., Vaswani, A., Uszkoreit, J., Kaiser, L., Shazeer, N., Ku, A.,
\& Tran, D. (2018). Image transformer. \emph{International Conference on
Machine Learning}, 4055--4064.

\leavevmode\hypertarget{ref-musdb18-hq}{}%
Rafii, Z., Liutkus, A., Stöter, F.-R., Mimilakis, S. I., \& Bittner, R.
(2019). \emph{MUSDB18-HQ - an uncompressed version of MUSDB18}.
\url{https://doi.org/10.5281/zenodo.3338373}

\leavevmode\hypertarget{ref-musdb18}{}%
Rafii, Z., Liutkus, A., Stöter, F.-R., Mimilakis, S. I., \& Bittner, R.
(2017). \emph{The {MUSDB18} corpus for music separation}.
\url{https://doi.org/10.5281/zenodo.1117372}

\leavevmode\hypertarget{ref-sawata20}{}%
Sawata, R., Uhlich, S., Takahashi, S., \& Mitsufuji, Y. (2020).
\emph{All for one and one for all: Improving music separation by
bridging networks}. \url{http://arxiv.org/abs/2010.04228}

\leavevmode\hypertarget{ref-stoter19}{}%
St "oter, F.-R., Uhlich, S., Liutkus, A., \& Mitsufuji, Y. (2019).
Open-unmix - a reference implementation for music source separation.
\emph{Journal of Open Source Software}.
\url{https://doi.org/10.21105/joss.01667}

\leavevmode\hypertarget{ref-fabian_robert_stoter_2019_3376621}{}%
Stöter, F.-R., \& Liutkus, A. (2019). \emph{Museval 0.3.0} (Version
v0.3.0) {[}Computer software{]}. Zenodo.
\url{https://doi.org/10.5281/zenodo.3376621}

\leavevmode\hypertarget{ref-Takahashi2021CVPR}{}%
Takahashi, N., \& Mitsufuji, Y. (2021). Densely connected multidilated
convolutional networks for dense prediction tasks. \emph{Proc. CVPR}.

\leavevmode\hypertarget{ref-vincent2006performance}{}%
Vincent, E., Gribonval, R., \& Févotte, C. (2006). Performance
measurement in blind audio source separation. \emph{IEEE Transactions on
Audio, Speech, and Language Processing}, \emph{14}(4), 1462--1469.

\end{CSLReferences}


@misc{musdb18,
  author       = {Rafii, Zafar and
                  Liutkus, Antoine and
                  Fabian-Robert St{\"o}ter and
                  Mimilakis, Stylianos Ioannis and
                  Bittner, Rachel},
  title        = {The {MUSDB18} corpus for music separation},
  month        = dec,
  year         = 2017,
  doi          = {10.5281/zenodo.1117372},
  url          = {https://doi.org/10.5281/zenodo.1117372}
}
\end{document}